\documentclass{ws-procs9x6}            

\usepackage{eepic}
\usepackage{indentfirst}
\usepackage{mathrsfs}
\usepackage{fancyhdr}
\usepackage{ulem}
\usepackage{float}
\usepackage{graphicx}
\usepackage[
colorlinks=true, linkcolor=black, breaklinks=true, urlcolor=black, citecolor=black
]{hyperref}
\usepackage{epstopdf}
\usepackage{bm,bbm}

\usepackage{xcolor}

\usepackage{tabularx}
\usepackage{subfigure}
\usepackage{epsfig}
\usepackage{color}
\usepackage{slashed}

\newcommand{\be}{\begin{equation}}
\newcommand{\ee}{\end{equation}}
\renewcommand{\L}{\mathscr{L}}

\newcommand{\bra}{\langle}
\newcommand{\ket}{\rangle}

\newcommand{\GeV}{\,\text{GeV}}

\usepackage{relsize}
\def\babar{\mbox{\slshape B\kern-0.1em{\smaller A}\kern-0.1em
    B\kern-0.1em{\smaller A\kern-0.2em R}}}

\begin{document}
\title{Light-quark components analysis and the nature of the $Y(4260)$}

\author{Yun-Hua~Chen}

\address{School of Mathematics and Physics, University
of Science and Technology Beijing, Beijing 100083, China}

\author{Ling-Yun~Dai}

\address{School of Physics and Electronics, Hunan University, Changsha 410082, China}

\author{Feng-Kun Guo}

\address{CAS Key Laboratory of Theoretical Physics,
             Institute of Theoretical Physics, Chinese Academy of Sciences,
Beijing 100190, China\\
School of Physical Sciences, University of Chinese Academy of
             Sciences, Beijing 100049, China}

\author{Bastian Kubis}

\address{Helmholtz-Institut f\"ur Strahlen- und Kernphysik (Theorie) and\\
             Bethe Center for Theoretical Physics,
             Universit\"at Bonn,
             53115 Bonn, Germany}

\begin{abstract}
We study the processes $e^+ e^-
\to Y(4260) \to J/\psi \pi\pi(K\bar{K})$.
The strong final-state interactions, especially the coupled-channel ($\pi\pi$ and $K\bar{K}$) final-state interaction in
the $S$-wave are taken into account in a model-independent way using dispersion
theory. It is found that the light-quark SU(3) octet state plays a significant
role in these transitions, implying that the $Y(4260)$ contains a
large light-quark component.
Our findings suggest that the $Y(4260)$ is neither a hybrid nor a conventional charmonium state. Furthermore, through an analysis of the ratio
of the light-quark SU(3) octet and singlet
components, we show that the $Y(4260)$ does not behave like a pure $\bar D D_1$ hadronic
molecule as well.
\end{abstract}


\vspace{8mm}

\bodymatter


The nature of the $Y(4260)$ has remained controversial
since its discovery by the \babar~Collaboration in the $J/\psi\pi^+\pi^-$ channel in the
initial-state radiation process $e^+e^-\to \gamma_{\rm ISR} J/\psi\pi^+\pi^-$. Its mass is inconsistent with
the naive quark model prediction for a normal vector charmonium,\cite{Godfrey} and it does not show
strong couplings to ground-state open-charm decay modes although it is above
the $D\bar{D}$ threshold. Such peculiar properties have initiated a
lot of theoretical and experimental studies, see Refs.~\citenum{Chen:2016qju,Guo:2017jvc} for recent reviews. In this work, we study the $e^+e^-\to Y(4260) \to J/\psi \pi\pi (J/\psi K\bar{K})$ processes to extract information on possible light-quark components of the $Y(4260)$. If the $Y(4260)$ contains no light quarks (as in the hybrid-state or the conventional-charmonium
scenarios), the light-quark source provided by the $Y(4260)$ has
to be in the form of an SU(3) flavor singlet state. Thus the study of the different light-quark SU(3) flavor eigenstate components provided by
the $Y(4260)$ in these processes may help to reveal its structure.

For the $Y(4260)J/\psi\pi\pi$ and
$Y(4260)J/\psi K\bar{K}$ contact couplings, the leading chiral effective Lagrangian
reads\cite{Mannel,Chen2016,Chen:2016mjn,Chen:2019gty}
\begin{align}\label{LagrangianYpsipipi}
\L_{Y\psi\Phi\Phi} &= g_1\bra V_{1}^\alpha J^\dag_\alpha \ket \bra u_\mu
u^\mu\ket +h_1\bra V_{1}^{\alpha} J^\dag_\alpha \ket \bra u_\mu u_\nu\ket
v^\mu v^\nu \notag\\&+g_8\bra  J^\dag_\alpha \ket \bra V_{8}^{\alpha} u_\mu u^\mu\ket
+h_8\bra J^\dag_\alpha \ket \bra V_{8}^{\alpha} u_\mu u_\nu\ket v^\mu v^\nu
+\mathrm{H.c.}\,,
\end{align}
where the parameters $g_1$ and $h_1$
correspond to the contributions from the SU(3) singlet component of the $Y(4260)$, and $g_8$ and $h_8$ are the corresponding parameters for the SU(3) octet component. The strong pion--pion final-state interactions (FSIs) as well as the $K\bar{K}$ coupled channel in the $S$-wave are taken into account model independently by using dispersion theory.
In the dispersion theory, the left-hand-cut contributions are approximated
by the sum of the $Z_c(3900)$-exchange mechanism and the triangle diagrams $Y(4260) \rightarrow \bar{D}D_1(2420)\rightarrow \bar{D}D^\ast\pi (\bar{D}D_s^\ast K )\rightarrow J/\psi\pi\pi(J/\psi K\bar{K})$.\cite{Cleven:2013mka,Albaladejo:2015lob} The
subtraction terms can be determined by matching to the chiral contact terms, since at low energies the amplitude should agree with the results required by chiral symmetry. To estimate the uncertainty due to the
dispersive input for the $\pi\pi/K\bar{K}$ rescattering, we will use two different $T_0^0(s)$ matrices, the
Dai--Pennington (DP)\cite{Dai:2014lza,Dai:2016ytz} and the Bern/Orsay (BO)\cite{Leutwyler2012,Moussallam2004} parametrizations, and compare the results. For details of the theoretical treatment, we refer to Ref.~\citenum{Chen:2019mgp}.

We fit to the experimental data of the $\pi\pi$
invariant mass spectra of $e^+e^- \to J/\psi
\pi^+\pi^-$ and the ratios of the cross sections ${\sigma(e^+e^- \to J/\psi
K^+ K^-)}/{\sigma(e^+e^- \to J/\psi \pi^+\pi^-)}$ measured at $E=4.26\GeV$ by the BESIII
Collaboration.\cite{Collaboration:2017njt,Ablikim:2018epj}
\begin{figure}[t]
\centering
\includegraphics[width=\linewidth]{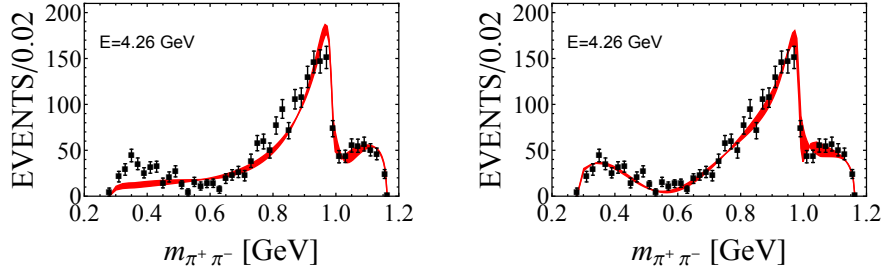}
\caption{Fit results of the $\pi\pi$ invariant mass spectra in
$e^+e^- \to J/\psi \pi^+\pi^-$ for Fit~a (left) and Fit~b (right).
The borders of the bands represent our best fit results using two different $T_0^0(s)$ matrices.
The experimental data are taken from Ref.~\citenum{Collaboration:2017njt}.
}\label{fig.Mpipi}
\end{figure}
To illustrate the effect of the SU(3) octet
component, we perform two fits. In Fit~a
we only consider the SU(3) singlet component, the
$Z_c$-exchange terms, and the triangle diagrams, while in Fit~b,
the SU(3) octet component is taken into account in addition.
The best fit results of the $\pi\pi$ mass spectra in $e^+e^- \to J/\psi \pi^+\pi^-$
are shown in Fig.~\ref{fig.Mpipi}. The fit results of the ratios of the cross sections ${\sigma(e^+e^- \to J/\psi K^+ K^-)}/{\sigma(e^+e^- \to J/\psi \pi^+\pi^-)}$ are given in Table~\ref{table-ratioes}.

\begin{table}[b]
\tbl{Experimental and theoretical values for the cross sections ratio ${\sigma(e^+e^- \to J/\psi K^+ K^-)}/{\sigma(e^+e^- \to J/\psi\pi^+\pi^-)}\times 10^{2}$ at $E=4.26\GeV$.}
{\begin{tabular}{ccccc}\toprule
Experiment &           Fit~a, DP &   Fit~b, DP&   Fit~a, BO &   Fit~b, BO   \\
\hline
 $4.99\pm 1.10$ & $4.46\pm 0.82$& $4.67\pm 0.98$& $5.37\pm 1.03$& $5.38\pm 0.82$   \\
\botrule
\end{tabular}
}
\label{table-ratioes}
\end{table}

One observes that Fit~a, in which the
SU(3) octet chiral contact terms are not included, cannot describes the experimental data well, especially for the broad peak below $0.6\GeV$.
In contrast, in Fit~b with the SU(3) octet chiral
contact terms added, the fit quality is improved significantly. Using the fit results, we can analyze the ratio of the parameters for the SU(3) octet component relative to those for
the SU(3) singlet component. In the DP parametrization we find $g_8/g_1=1.2 \pm 0.2$ and $h_8/h_1=57\pm 76$, while in the BO parametrization we obtain $g_8/g_1=1.1 \pm 0.1$ and $h_8/h_1=102\pm 152$,
which agree well with each other within errors.
Note that in the $\bar{D}D_1$
hadronic molecule scenario of the $Y(4260)$, the light-quark component is in the form of $|u \bar{u}+d\bar{d}\rangle/\sqrt{2} = (\sqrt{2}
V_1^{\text{light}} +V_8^{\text{light}})/\sqrt{3}$, where $V_1^{\text{light}}=
\frac{1}{\sqrt{3}}(u\bar{u}+d\bar{d}+s\bar{s})$ and $
V_8^{\text{light}}=\frac{1}{\sqrt{6}}(u\bar{u}+d\bar{d}-2
s\bar{s})$.
They give the ratio of $1/\sqrt{2}$, which differs significantly from our results (given by the values of $g_8/g_1$).
Therefore we conclude that the $Y(4260)$ contains a large
light-quark SU(3) octet component, and the
scenarios of a hybrid or conventional charmonium are disfavored
since the light quarks have to be produced in an SU(3)
singlet configuration inside such states. Also our study shows that the $Y(4260)$ cannot be a pure $\bar{D}D_1$ hadronic molecule as well.

\begin{figure}[t]
\centering
\includegraphics[width=\linewidth]{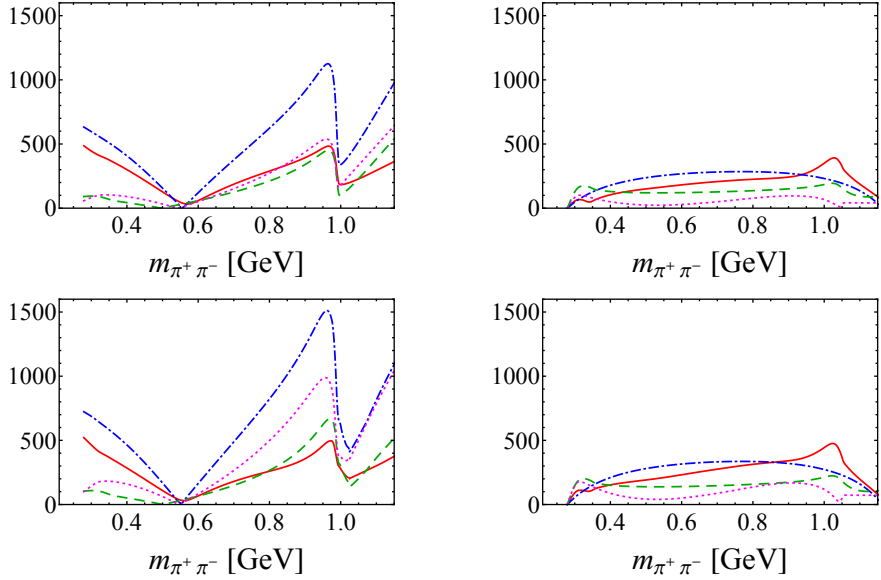}
\caption{The moduli of the $S$- (left) and $D$-wave (right)
amplitudes for $e^+e^- \to J/\psi \pi^+\pi^-$ in
Fit~b, using the DP (top) or the BO (bottom) parametrizations. The red solid lines represent our best fit results,
while the blue dot-dashed, darker green dashed,
and magenta dotted lines correspond to the contributions
from the chiral contact terms, $Z_c$-exchange, and the triangle diagrams, respectively.
}
\label{fig.Moduli}
\end{figure}
In Fig.~\ref{fig.Moduli}, we plot the moduli of the $S$- and
$D$-wave amplitudes from the chiral contact terms, the $Z_c$-exchange terms, and the triangle diagrams for Fit~b.
One observes that the $D$-wave contribution is comparable
to the $S$-wave contribution in almost the whole energy region. Such a large $D$-wave
contribution again indicates that the $Y(4260)$ cannot be a conventional charmonium state, for which the $\pi\pi$ $S$-wave should be dominant.
Also note that in the $\bar D D_1$ hadronic molecule interpretation,\cite{Cleven:2013mka,Lu:2017yhl} the $\pi\pi$ $D$-wave emerges naturally since the $D_1$ decays dominantly into $D$-wave $D^*\pi$.

At last, we anticipate a combined analysis of both the $Y(4260)$ and $Z_c(3900)$ data, which will be crucial to reveal the nature of both states.

\section*{Acknowledgments}

This work is supported in part by the Fundamental Research Funds
for the Central Universities under Grants No.~531107051122 and No.~FRF-BR-19-001A, by the National Natural Science Foundation of China (NSFC) under Grants Nos.~11975028, 11805059, 11847612, and No.~11835015, by NSFC and Deutsche Forschungsgemeinschaft (DFG) through
funds provided to the Sino--German Collaborative Research Center ``Symmetries and the
Emergence of Structure in QCD'' (NSFC Grant No.~11621131001,
DFG Grant No.~TRR110), by the Thousand Talents Plan for Young
Professionals, by the  Chinese Academy of Sciences (CAS) Key Research Program of Frontier Sciences
(Grant No.~QYZDB-SSW-SYS013), by the CAS Key Research Program (Grant No.~XDPB09), and by the CAS Center for Excellence in Particle Physics (CCEPP).

\bibliographystyle{ws-procs9x6} 
\bibliography{ws-pro-sample}

\end{document}